\documentclass[12pt]{article}
\usepackage{graphicx}

\newcommand\pubnumber{SNSN-323-63}
\newcommand\pubdate{\today}

\def\aarhus{Department of Physics and Astronomy, Aarhus University, Ny Munkegade, DK-8000 Aarhus C, Denmark}

\def\Title#1{\begin{center} {\Large #1 } \end{center}}
\def\Author#1{\begin{center}{ \sc #1} \end{center}}
\def\Address#1{\begin{center}{ \it #1} \end{center}}

\newcommand\pubblock{\rightline{\begin{tabular}{l} \pubnumber\\
         \pubdate  \end{tabular}}}
\newenvironment{Abstract}{\begin{quotation}  }{\end{quotation}}
\newenvironment{Presented}{\begin{quotation} \begin{center} 
             PRESENTED AT\end{center}\bigskip 
      \begin{center}\begin{large}}{\end{large}\end{center} \end{quotation}}

\def\beq{\begin{equation}}
\def\eeq#1{\label{#1}\end{equation}}
\def\eeqn{\end{equation}}

\def\beqa{\begin{eqnarray}}
\def\eeqa#1{\label{#1}\end{eqnarray}}
\def\eeqan{\end{eqnarray}}

\let\bar=\overbar

\def\Dslash{\not{\hbox{\kern-4pt $D$}}}
\def\dslash{\not{\hbox{\kern-2pt $\del$}}}

\def\msb{{\bar{\ssstyle M \kern -1pt S}}}

\begin{document}
\begin{titlepage}
\pubblock

\vfill
\Title{Neutrino physics and precision cosmology}
\vfill
\Author{Steen Hannestad}
\Address{\aarhus}
\vfill
\begin{Abstract}
I review the current status of structure formation bounds on neutrino properties such as mass and energy density. I also discuss future cosmological bounds as well as a variety of different scenarios for reconciling cosmology with the presence of light sterile neutrinos.
\end{Abstract}
\vfill
\begin{Presented}
 NuPhys2015, Prospects in Neutrino Physics\\
Barbican Centre, London, UK,  December 16--18, 2015
\end{Presented}
\vfill
\end{titlepage}
\def\thefootnote{\fnsymbol{footnote}}
\setcounter{footnote}{0}

\section{Introduction}

One of the most fruitful lines of research in astroparticle physics has been the study of how cosmological structure
formation can be used to probe neutrino physics (see \cite{Lesgourgues:2006nd,Hannestad:2010kz,Wong:2011ip} for more details).
Over the past decade observations of the Cosmic Microwave Background and the large scale distribution of galaxies have
made it possible to put a bound on the mass of standard model neutrinos which is well below 0.5 eV for the sum of neutrino masses \cite{Ade:2015xua}, $\sum m_\nu$, {\it provided} that the $\Lambda$CDM model is basically the correct cosmological model {\it and} that neutrino physics is described by 
standard model physics.
Depending on the specific combination of data sets used the current upper bound can be as low as 0.12-0.13 eV (see e.g.\ \cite{Palanque-Delabrouille:2015pga,Cuesta:2015iho}), close to the mass predicted in the inverted hierarchy for a massless $\nu_3$.  

Cosmology therefore seems to be at the brink of a neutrino mass detection.
However, more complex models of both cosmology and neutrino physics can change this picture and significantly shift the mass bound or even
evade it altogether. The purpose of this short review is to discuss the status of cosmological neutrino measurements, including the 
caveats in bounds on neutrino properties. I will also discuss the possibility of probing neutrino physics beyond the standard model in the 
form of additional sterile neutrino species and new non-standard interactions.

In the standard model neutrino interactions at low energies are well described by Fermi theory. In the universe at temperatures well below the QCD phase transition neutrinos interact dominantly with electrons, positrons, and other neutrinos.
Under the assumption that chemical or pseudo-chemical potentials are small for all the involved species the interaction rates can be calculated very precisely and the process of neutrino freeze-out followed with adequate precision.
The outcome is that the neutrino distribution is well described by a thermal distribution of temperature $T_\nu \sim \left(\frac{4}{11}\right)^{1/3} T_\gamma$, with corrections from incomplete neutrino decoupling at $e^+ e^-$ annihilation and finite temperature QED effects entering at the 1\% level \cite{Mangano:2005cc} (see \cite{Grohs:2015tfy} for a recent discussion).
At late times the main impact of neutrinos on structure formation can be quantified by their contribution to the energy density. Since the neutrino contribution in the late time universe is dominated by the rest mass term and the number density can be calculated exactly, the only unknown quantity is the sum of neutrino masses, $\sum m_\nu$. This is related to the physical energy density through
\begin{equation}
\Omega_\nu h^2 \simeq \frac{\sum m_\nu}{94.6 \, {\rm eV}}.
\end{equation}
This simple relation means that if neutrino physics is describable in terms of pure standard model physics the relation of the cosmological parameter $\Omega_\nu h^2$ to the fundamental particle physics parameter $\sum m_\nu$ is trivial. However, as will be discussed later this simple relation 
does not hold for a wide variety of beyond the standard model scenarios.

\section{Large scale structure observables and neutrinos}

Even standard model neutrinos are a source of dark matter because of their finite, but non-zero masses. However,
neutrinos impact cosmological structure formation in a unique way because they are much lighter than other particles contributing to the matter density.
The fact that neutrinos are very light and weakly interacting means that they can stream over large distances, erasing any pre-existing structure in the neutrino density. If treated as a fluid this is equivalent to having a significant anisotropic stress component.
This very significantly suppressed fluctuation power in neutrinos on all subhorizon scales. In the CMB this can in turn also be seen as a suppression of power because the neutrino contribution to the metric source term in the Boltzmann equation for photons is suppressed.

For typical sub-eV neutrino neutrino masses, neutrinos are still relativistic during recombination and therefore the direct effect of a non-zero neutrino mass is limited in the primary CMB signal. However, non-zero neutrino masses strongly influence the subsequent growth of perturbations
and lead to suppression of power on all scales below the free-streaming scale. In the case of the CMB power spectrum this can be seen as a reduction in the effect of weak gravitational lensing on the CMB spectrum.

In the large scale structure power spectrum the effect is even more pronounced because the matter power spectrum (rather than the photon power spectrum) is probed directly. The matter power spectrum is suppressed by a large factor (roughly $\delta P/P \sim -8 f_\nu$, where $f_\nu=\Omega_\nu/\Omega_m$) on all scales smaller than the free-streaming scale \cite{Lesgourgues:2006nd}, and even in the most extreme case of a massless $\nu_1$ and the normal hierarchy where $\sum m_\nu \sim 0.06$ eV the effect on the power spectrum is several percent. This is close to the precision with which the matter power spectrum can currently be measured. However, even the largest current surveys are not able to probe distances comparable to the free-streaming length. The consequence is that the effect of neutrinos is simply seen as an overall and scale-independent reduction in power which is difficult to disentangle from other effects.

\section{Current and future bounds}

One of the most important developments in the last decade has been that cosmological data is now of sufficient quality that strong bounds on neutrino masses can be obtained using at most two different types of data with well-understood systematics.
Using the CMB data from Planck alone provides a bound of 0.59 eV in the case of the $\Lambda$CDM model \cite{Ade:2015xua}. When large scale structure data is added the bound is strengthened to 0.23 eV \cite{Ade:2015xua}. Further improvements can be made by adding e.g.\ data from the Lyman-$\alpha$ forest, although this also means that the bound becomes less robust.
We note that the current upper bound on the sum of neutrino masses is getting relatively close to the predicted mass for a massless $\nu_1$ in the inverted hierarchy, but that at present it is impossible to use cosmological data to distinguish the two hierarchies.

\begin{figure}[h]
\includegraphics[width=8.0cm,angle=270]{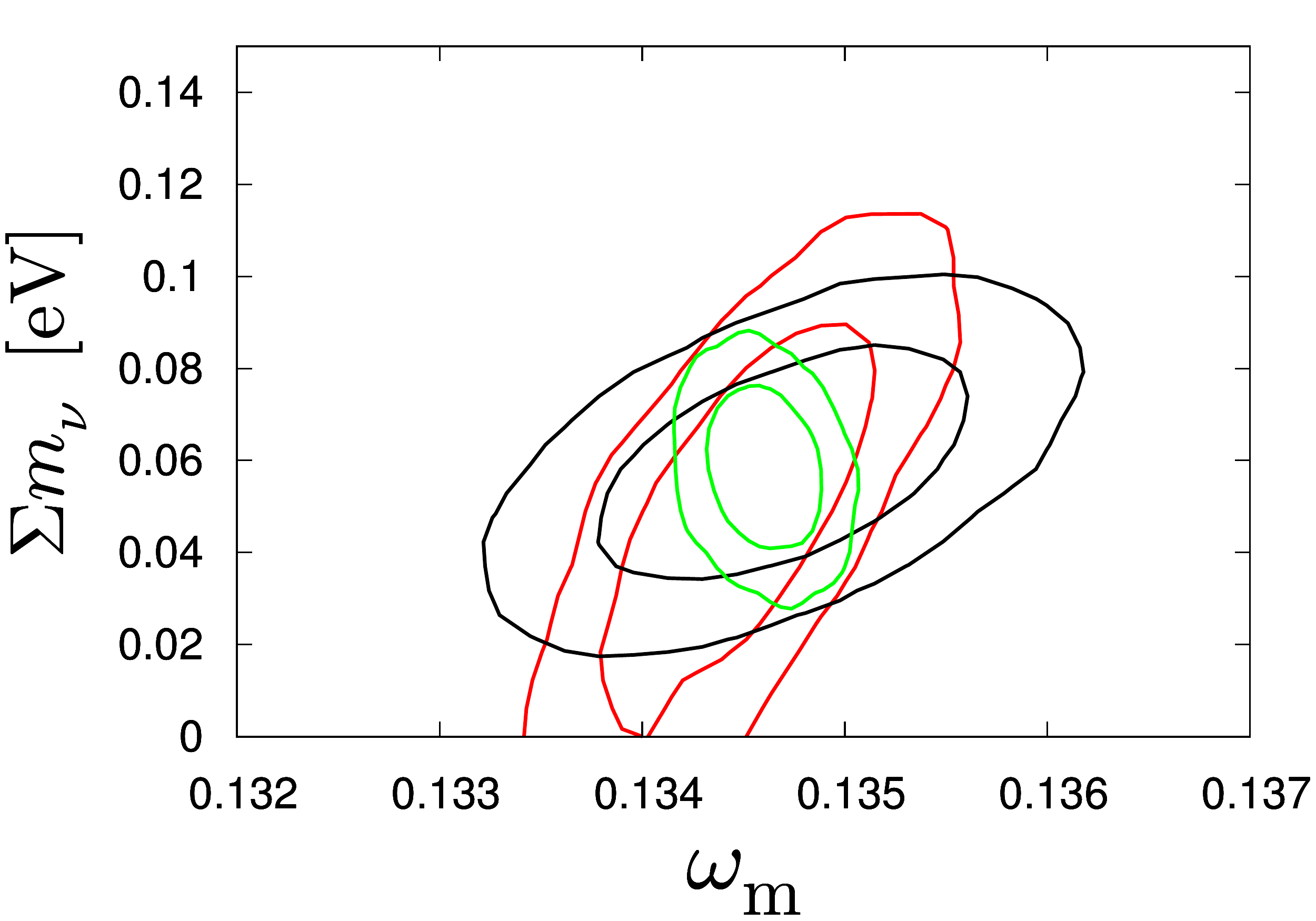}
\caption{
68\% and 95\% constraints on the matter density, $\omega_m=\Omega_m h^2$, and the sum of neutrino
masses,
$\sum m_\nu$, for the combination of Planck CMB data with
EUCLID shear data (red curves)
and EUCLID galaxy data (black curves), as well as the combination of all three (green curves).
The data used corresponds to the ``csgx'' case in \cite{Hamann:2012fe}.
\label{fig:abs}}
\end{figure}

The coming decade is likely to see as big an improvement on neutrino constraints as the previous. 
In particular the large scale structure data from very large surveys such as EUCLID \cite{Laureijs:2011gra} will improve neutrino constraints radically. Ref.~\cite{Abazajian:2013oma} provides a good overview of projected sensitivities from data sets becoming available within the next decade (see also \cite{Allison:2015qca}). Combining the Planck data with just one of these observables in many cases leads to a sensitivity good enough to make a neutrino mass detection very likely. When several large scale structure observables (lensing, galaxy power spectrum, cluster mass function) are combined with the Planck data the formal sensitivity might approach the 10 meV level \cite{Hamann:2012fe,Basse:2013zua,Basse:2014qqa}, indicating that a 5$\sigma$ detection of the neutrino mass might be feasible.
Fig.~\ref{fig:abs} (taken from \cite{Hamann:2012fe}) shows projected confidence contours from combining a Planck-like CMB experiment with data from EUCLID.

Of course it should be kept in mind that even when a hot dark matter component is unambiguously detected it will not be possible to state that it consists of neutrinos. Since structure formation is only sensitive to the kinematical properties of the hot dark matter and not to its actual particle content a final confirmation has to await the measurement of neutrino masses in e.g.\ tritium decay experiments such as KATRIN \cite{Osipowicz:2001sq}.

\section{Non-standard neutrino physics}

Within the standard model the only unknown parameter relevant to cosmological structure formation is the sum of neutrino masses. However, types of beyond standard model physics predict the presence of one or more non-standard features in the neutrino sector.
From a cosmological point of view the simplest possibility is that the neutrino number density (and thus the contribution to the energy density while neutrinos are still relativistic) is different from what is predicted in the standard model.
The standard parameter used in cosmological parameter fitting is $N_{\rm eff}$, defined as
\begin{equation}
N_{\rm eff} \equiv \frac{\rho_\nu}{\rho_{\nu,0}}
\end{equation}
with $\rho_{\nu,0} = \frac{7}{8} \left(\frac{4}{11}\right)^{4/3} \rho_\gamma$, measured at $T \gg m_\nu$, i.e.\ it corresponds to the energy density in relativistic neutrinos at early times when they are relativistic. The standard model prediction is $N_{\rm eff} \simeq 3.046$, but a variety of beyond the standard model scenarios could lead to a a higher (or lower) value of $N_{\rm eff}$. 

We note that using $\sum m_\nu$ and $N_{\rm eff}$ is typically not enough to fully specify a given model because $N_{\rm eff}$ cannot be uniquely translated into the number density and therefore the current neutrino density, $\Omega_\nu h^2$, cannot be calculated using these two parameters alone.
However, in scenarios where the particle distribution is reasonably close to thermal $N_{\rm eff}$ and $\sum m_\nu$ can be used as fitting parameters.

The current Planck data shows no indication in itself that $N_{\rm eff}$ is different from the value predicted by the standard model \cite{Ade:2015xua}. However, it should be noted that the Planck inferred value of $H_0$ is significantly different from the one measured directly in the local Universe
\cite{Riess:2016jrr},
and that a possible way of lessening the tension between the two results is to have a significantly higher $N_{\rm eff}$ than predicted by the standard model.

Another important aspect of neutrino physics which can be tested by cosmology is the presence of non-standard interactions. Well-known possibilities are additional massive vector bosons or light pseudoscalars or scalars coupling to neutrinos. Such interactions are relatively hard to probe in direct experiments and cosmology can provide very non-trivial bounds (see e.g.\ \cite{Archidiacono:2013dua} for a recent discussion). 

Finally, there are several indications from terrestrial neutrino experiments that additional sterile neutrinos might be present (see e.g.\ \cite{Abazajian:2012ys} for an overview) .
However, such neutrinos are extremely disfavoured by cosmological data. The reason is that the relatively large mixing angle required to explain oscillation data inevitably leads to almost complete thermalisation of the additional mass states. Therefore the additional neutrinos act essentially like active neutrinos with the same mass and since the preferred mass is around 1 eV they are disfavoured by more than 5$\sigma$ by the usual cosmological mass bound.

If the sterile neutrino hypothesis is confirmed an additional modification of either cosmology or neutrino physics is required.
One possible solution is to add a large neutrino chemical potential. This has the effect of delaying the onset of the resonant production of sterile neutrinos until after the active neutrinos decouple and leads to a lower abundance of sterile neutrinos. However, a more attractive scenario is to add additional interactions in the sterile sector. This leads to a self-generated matter potential which also has the effect of delaying thermalisation of the sterile state.

Two basic scenarios have been studied so far. The first is based on a new light, but not massless vector boson mediating a Fermi-like interaction between the sterile states
\cite{Hannestad:2013ana,Dasgupta:2013zpn,Bringmann:2013vra,Saviano:2014esa,Mirizzi:2014ama,Tang:2014yla,Chu:2015ipa}. The new interaction can be significantly stronger than the ordinary weak interaction without violating any known constraints, and can delay sterile neutrino thermalisation until long after active neutrino decoupling. However, in most cases the sterile state will thermalise at low temperature and thus the cosmological neutrino mass constraint still poses a potential problem for this scenario \cite{Mirizzi:2014ama,Chu:2015ipa}.

\begin{figure}[h]
\includegraphics[width=8.0cm]{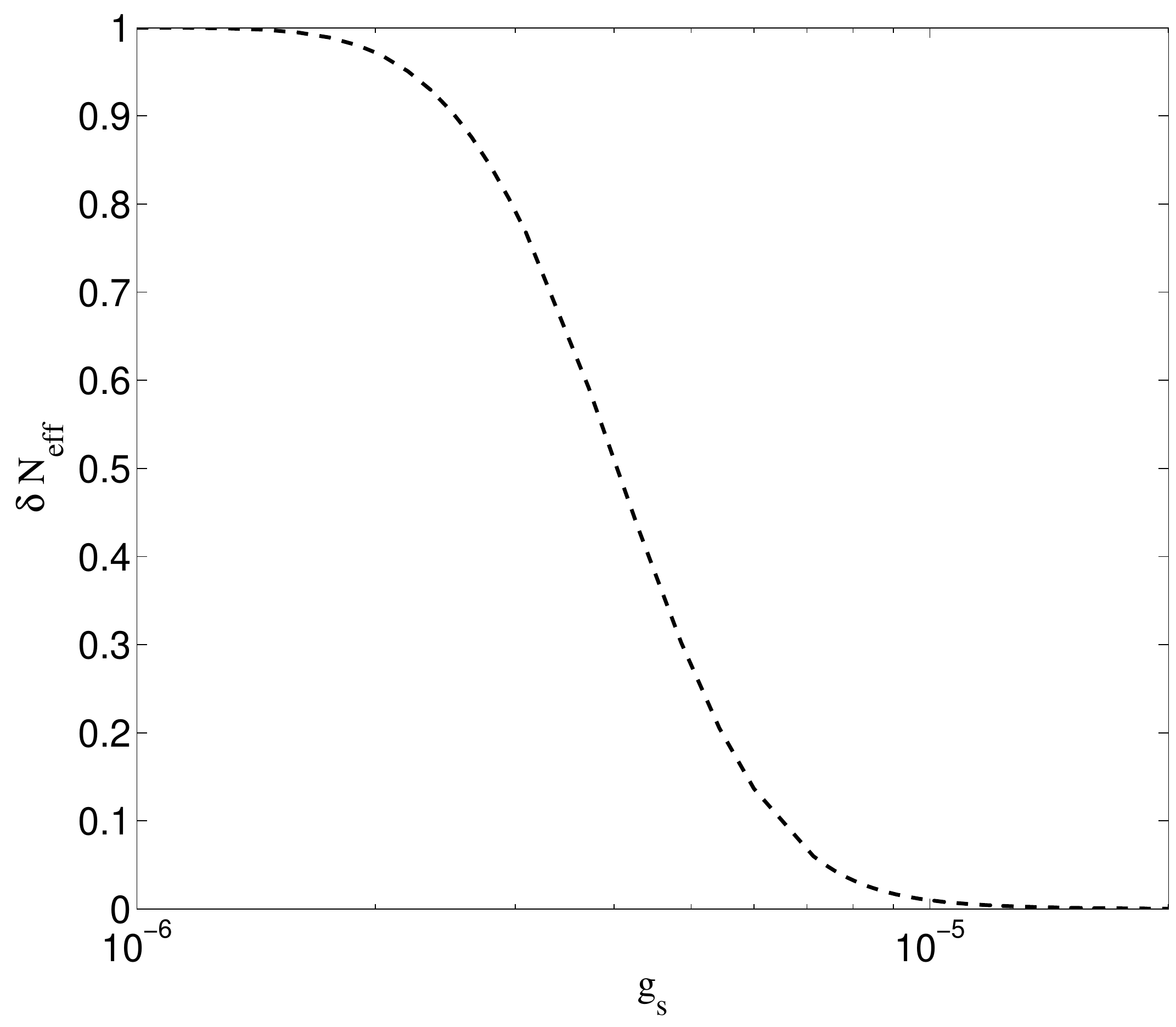}
\caption{The contribution of the sterile neutrino to the relativistic energy density $\delta N_{\rm eff} = N_{\rm eff} - 3$ as a function of the dimensionless coupling parameter $g_s$. \label{fig:abs2}}
\end{figure}

Another scenario is based on a new massless pseudoscalar or scalar \cite{Archidiacono:2014nda,Archidiacono:2015oma}. Fig.~\ref{fig:abs2} shows how the sterile neutrino contribution to $N_{\rm eff}$ can be reduced when the dimensionless coupling strength, $g_s$ is increased (taken from \cite{Archidiacono:2014}). While similar to the vector mediator scenario at high temperatures it is qualitatively very different at low temperatures because the interaction becomes stronger at low temperatures so that the sterile neutrino and scalars/pseudoscalars form a strongly interacting fluid at low temperatures. This setup is very similar to the ``neutrinoless universe'' scenario previously studied for active neutrinos in which the active neutrinos couple to a new light particle and disappear through pair annihilation as soon as they become non-relativistic, thereby alleviating the effect of the mass on the matter power spectrum \cite{Beacom:2004yd}. This scenario has been ruled out by CMB observations because it would lead to an excessive enhancement of fluctuations on sub-horizon scales prior to recombination \cite{Hannestad:2004qu}.
However, provided that only the mainly sterile mass state couples to the new particle the scenario can actually provide a better fit to current data than $\Lambda$CDM, and interestingly it strongly prefers a value of the Hubble parameter close to the locally measured value \cite{Archidiacono:2015oma}.

\section{Discussion}

I have reviewed the current status of light neutrinos in cosmological structure formation. Within the class of $\Lambda$CDM-like models with upper bound on the sum of neutrino masses is now robustly below 0.5 eV. Depending on assumptions about the cosmological model and the data used the bound might even be close to 0.1 eV, but this is a much less robust statement.

Within the next 5-8 years new data from a host of different large scale structure surveys will drastically improve the sensitivity to the neutrino mass. It can reasonably be expected that future data will bring the sensitivity down to the 10 meV level and allow for a detection of the neutrino hot dark matter component of the universe.

Cosmology can also be used as a sensitive probe of neutrino physics beyond the standard model. The prime example of this is the use of cosmology to constrain the existence of sterile neutrinos.
There are currently several hints from terrestrial neutrino experiments that one or more additional sterile states of eV mass might be needed to explain data. Furthermore the additional mass states are required to have relatively large mixing with the active neutrinos. This in turn leads to almost complete thermalisation of the new state in the early Universe and the bound derived for active neutrinos apply, disfavouring such additional sterile states at 5$\sigma$ or more.

If the sterile hypothesis is confirmed it will therefore require modifications to either cosmology or neutrino physics in order to be reconciled with cosmological data. One possibility which is currently receiving significant attention is the possibility that sterile neutrinos possess new interactions. Such interactions can be quite strong without violating any current experimental or astrophysical constraints and can reconcile eV sterile neutrinos with cosmology. In fact, interactions via a massless mediator which makes sterile neutrinos strongly interacting at late times can provide a better fit to data than $\Lambda$CDM in terms of $\chi^2$.

In conclusion, cosmology is an extremely powerful laboratory for probing neutrino physics. This will remain true in the coming decade and cosmology will almost certainly provide the first detection of the neutrino hot dark matter component of the universe in this timespan.

\end{document}